\documentclass[preprint2,times,tighten]{aastex6}
\usepackage{graphicx,amssymb,aas_macros}
\usepackage{amsmath} 
\usepackage{hyperref} 
\usepackage{amssymb} 
\usepackage{graphicx} 
\usepackage{indentfirst} 

\newcommand{\apg}{\:^{>}_{\sim}\:}

\newcommand{\cmjj}{\mbox{${\rm cm^{-2}}$}}
\newcommand{\cmjjj}{\mbox{${\rm cm^{-3}}$}}
\newcommand{\etal}{et al.}

\newcommand{\kms}{\mbox{km\ s${^{-1}}$}}

\shorttitle{Multiphase gas in a massive elliptical galaxy at $z=0.4$}
\shortauthors{Zahedy et al.}

\begin{document}


%%\title[Molecular hydrogen in a massive elliptical at $z=0.4$]{Discovery of warm molecular hydrogen in a massive elliptical galaxy at $\lowercase{z}=0.4$} 

%\title[]{Multiphase Gaseous Environment of Quenched Galaxies: Discovery of $\mathrm{H_2}$ and Highly Ionized Oxygen within a Massive Elliptical at $\lowercase{z}=0.4$} 
%\title[]{Constraining Feedback with the Discovery of Multiphase Gas in a Massive Elliptical at $\lowercase{z}=0.4$} 
\title[]{Evidence for Late-Time Feedback from the Discovery of Multiphase Gas in a Massive Elliptical at $\lowercase{z}=0.4$} 
%\title[]{Discovery of Multiphase Gas in a Massive Elliptical at $\lowercase{z}=0.4$: Evidence for Late-time Feedback} 

%% Use \author, \affil, and the \and command to format author and affiliation 
%% information.  If done correctly the peer review system will be able to
%% automatically put the author and affiliation information from the manuscript
%% and save the corresponding author the trouble of entering it by hand.
%%
%% The \affil should be used to document primary affiliations and the
%% \altaffil should be used for secondary affiliations, titles, or email.

%% Authors with the same affiliation can be grouped in a single

\author{Fakhri S. Zahedy\altaffilmark{1}, 
Hsiao-Wen Chen\altaffilmark{2},
Erin Boettcher\altaffilmark{2},
Michael Rauch\altaffilmark{1}, 
K. Decker French\altaffilmark{3,1},
and Ann I. Zabludoff\altaffilmark{4}}

\altaffiltext{1}{The Observatories of the Carnegie Institution for Science, Pasadena, CA 91101, USA; \href{mailto:fzahedy@carnegiescience.edu}{fzahedy@carnegiescience.edu}}
\altaffiltext{2}{Department of Astronomy \& Astrophysics, The University of Chicago, Chicago, IL 60637, USA} 
\altaffiltext{3}{Department of Astronomy, University of Illinois Urbana-Champaign, Urbana, IL 61801, USA}
\altaffiltext{4}{Department of Astronomy \& Steward Observatory, University of Arizona, Tucson, AZ 85721, USA}


%% AASTeX 6.0 supports the ability to suppress the names and affiliations
%% of some authors and displaying them under a "collaboration" banner to
%% minimize the amount of author information that to be printed.  This 
%% should be reserved for articles with an extreme number of authors.  
%% The necessary command are \AuthorCallLimit and \collaborationName.
%% An \AuthorCallLimit=2 call prior to the author list will only show
%% the authors in the first two \author calls.  The \collaborationName
%% defines the collaboration identifier.  Commented examples below.

%\AuthorCallLimit=1
%% Will only show Schwarz & Muench since Schwarz and Muench
%% are in the same \author call. 
%\collaborationName{Friends of AASTeX}
%% will print "The AAS collaboration" after the shortened author list.
%% Note that all the \altaffil information will still be shown so it
%% has to be manually commented out if you do not want it shown.
%%
%% Note that all of these author will be shown in the published article.
%% This feature is meant to be used prior to acceptance to make the
%% front end of a long author article more manageable.

\begin{abstract}

%Characterizing the gaseous properties of quiescent galaxies is critical to understanding the role of quenching in galaxy evolution. 
We report the first detection of multiphase gas within a quiescent galaxy beyond $z\approx0$. The observations 
use the brighter image of doubly lensed QSO HE\,0047$-$1756 to probe the ISM of the massive 
($M_{\rm star}\approx 10^{11}\, \mathrm{M_\odot}$) elliptical lens galaxy at $z_\mathrm{gal}=0.408$. 
Using {\it Hubble Space Telescope}'s Cosmic Origins Spectrograph (COS), we obtain a medium-resolution FUV spectrum of the
lensed QSO and identify numerous absorption features from $\mathrm{H_2}$
in the lens ISM at projected distance $d=4.6$ kpc. The $\mathrm{H_2}$ column density
is $\log\,N(\mathrm{H_2})/\cmjj=17.8^{+0.1}_{-0.3}$ with a molecular gas fraction of  $f_\mathrm{H_2}=2-5\%$, roughly consistent with some local quiescent galaxies. 
The new COS spectrum also reveals kinematically complex absorption features from highly ionized species \ion{O}{6} and \ion{N}{5} with column densities log $N$(\ion{O}{6})$/\cmjj =15.2\pm0.1$ and  log\, $N$(\ion{N}{5})$/\cmjj\ =14.6\pm0.1$, among the highest known in external galaxies. 
Assuming the high-ionization absorption features originate in a transient warm ($T\sim10^5$ K) phase undergoing radiative cooling from a hot halo surrounding the galaxy, we infer a mass accretion rate of $\sim 0.5-1.5\,\mathrm{M_\odot\,yr^{-1}}$. The lack of star formation in the lens suggests the bulk of this flow is returned to the hot halo, implying a heating rate of $\sim10^{48}\,\mathrm{erg\,yr^{-1}}$. Continuous heating from evolved stellar populations (primarily SNe Ia but also winds from AGB stars) may suffice to prevent a large accumulation of cold gas in the ISM, even in the absence of strong feedback from an active nucleus.

\end{abstract}

%\keywords{galaxies: halos --- galaxies: elliptical and lenticular, cD ---
 % quasars: absorption lines  --- galaxies: abundances}



\section[]{Introduction} 

How and why some galaxies cease forming stars and remain quiescent are open questions that bear significantly on our understanding of galaxy evolution.
Contrary to the expectation that a lack of star-formation is the consequence of a paucity of cool gas, observational studies have established that a high fraction of passive galaxies are not gas-poor (see Chen 2017a and references therein). Systematic 21cm surveys have discovered that more than a third of present-day quiescent galaxies contain abundant neutral hydrogen (\ion{H}{1}) gas in their interstellar medium (ISM; e.g., Oosterloo \etal\ 2010; Serra \etal\ 2012). At an earlier epoch, QSO absorption-line surveys of \ion{Mg}{2} absorption features near luminous red galaxies (LRGs) at $z\sim0.5$ have also demonstrated that a significant fraction of these distant massive ellipticals (with total stellar masses of $M_\mathrm{star} \gtrsim 10^{11}\,\mathrm{M}_\odot$) are surrounded by chemically enriched cool gaseous halos on $\sim100$ kpc scales (e.g., Gauthier \etal\ 2009, 2010; Bowen \& Chelouche 2011; Huang \etal\ 2016; Chen \etal\ 2018). The total mass in this cool ($T\sim10^4$ K) circumgalactic medium (CGM) is estimated to be $M_\mathrm{cool} \approx (1-2)\times10^{10}\,\mathrm{M}_\odot$ within projected distance $d<160$ kpc (or as much as $\approx 4\times10^{10}\,\mathrm{M}_\odot$ at $d < 500$ kpc; Zahedy \etal\ 2019), similar to what has been reported for star-forming galaxies (e.g., Chen \etal\ 2010; Stocke \etal\ 2013; Werk \etal\ 2014).

The existence of large reservoirs of cool gas around massive ellipticals challenges simple theoretical expectations that these galaxies are surrounded by predominantly hot ($T\gtrsim10^6$ K) gas on both small ($\lesssim 10$ kpc; ISM) and large ($\sim 100$ kpc; CGM) scales. Furthermore, it indicates that some physical mechanisms are preventing the gas from triggering the resumption of star formation in the central galaxy. 
A common feature of the gaseous environment at $d\lesssim 10$ kpc around massive quiescent galaxies is the high $\mathrm{Fe/Mg}$ abundance ratio, $[\mathrm{Fe}/\mathrm{Mg}]\apg0$, that has been observed in every instance cool gas is present (Zahedy \etal\ 2016, hereafter Z16; Zahedy \etal\ 2017a). This Fe enhancement not only indicates that the ISM of massive ellipticals has been significantly enriched by Type Ia supernovae (SNe Ia), but also points to SNe Ia as a potentially important heating source in massive halos (e.g., Conroy \etal\ 2015; Li \etal\ 2020a,b). 

One of the galaxies studied in Z16 is a massive ($M_{\rm star}\approx 10^{11}\, \mathrm{M_\odot}$) elliptical lens for QSO HE0047$-$1756 at $z_\mathrm{gal}=0.408\pm0.001$. It exhibits extremely strong and kinematically complex low-ionization metal absorptions with a line-of-sight velocity spread exceeding 600 \kms\ (Figure 1, top) and a velocity shear of $\approx350$ \kms\ between two locations $\approx8$ kpc apart in projection. Long-slit far-ultraviolet (FUV) spectroscopic observations of both lensed QSO images revealed the presence of abundant \ion{H}{1} within the galaxy, with measured \ion{H}{1} column densities of log\,$N$(\ion{H}{1})$/\cmjj= 19.6-19.7$ at both locations (Zahedy \etal\ 2017b, hereafter Z17), constraining the gas metallicity to be $\mathrm{[Fe/H]}\gtrsim 0$ for both sightlines after accounting for likely dust depletion. While Z17 also noted the presence of possible absorption features from other metal ions probing a wide range of ionization states, including the highly ionized \ion{O}{6} $\lambda\lambda1031,1037$ doublet, their low-resolution spectra precluded a detailed investigation of these absorption profiles to confirm the presence of high ions. Because $\mathrm{O^{5+}}$ ions are most abundant at temperatures near the peak of the cooling curve for metal-enriched gas ($T\approx 10^{5.5}$\,K; e.g., Gnat \& Sternberg 2007), such warm gas is expected to cool rapidly if left to itself. Therefore, the possible detection of rapidly cooling gas in the ISM implies the presence of an effective heating mechanism in the galaxy. Characterizing the properties of such a transient gas phase and its relationship to cooler atomic/molecular gases offers a unique opportunity to understand the dynamic gas content of ellipticals, in order to gain insight into late-time feedback in massive quiescent galaxies.

In this {\it Letter}, we report the robust detection of highly ionized gas in the ISM of the massive elliptical lens of HE0047$-$1756, traced by the \ion{O}{6} and \ion{N}{5} absorption features. Furthermore, we report the serendipitous discovery of molecular hydrogen ($\mathrm{H_2}$) in the ISM, the first direct detection of $\mathrm{H_2}$ within a passive galaxy beyond the local Universe. We compare the spatial distributions and mass budgets of the molecular ($T\sim100$\,K), cool ($T\sim10^4$\,K), and warm ($T\sim10^5$\,K) ISM phases and discuss their implications for feedback in massive ellipticals. We adopt a $\Lambda$ cosmology with $\Omega_{\rm M}=0.3$, $\Omega_\Lambda = 0.7$, and $H_0 =70 \ {\rm km} \ {\rm s}^{-1}\ {\rm Mpc}^{-1}$.

\section[]{Observations}

New FUV spectra of image {\it A} of the doubly lensed QSO HE\,0047$-$1756 ($z_\mathrm{QSO} = 1.676$; Figure 1 of Z16) were obtained with the Cosmic Origins Spectrograph (COS) onboard the {\it Hubble Space Telescope (HST)} during our {\it HST}  Cycle 25 observing program (Program ID: 15250; PI: Zahedy) in December 2018. {\it HST}/COS with the G130M and G160M gratings provides a wavelength coverage from $\lambda\approx1130$ \AA\  to $\lambda \approx 1790$ \AA\ at a medium resolution of ${\rm FWHM}\approx18-20\, \kms$, a fifteenfold increase in resolution from the Z17 spectra. The total integration time of the observations was 9,418 s and 17,722 s for the G130M and G160M gratings, respectively, comprising 22 individual exposures spread over three separate {\it HST} visits. The observations used two (four) central wavelength settings for the G130M (G160M) grating and two or four FP-POS at each central wavelength, to ensure a continuous wavelength coverage and reduce fixed pattern noise over the full spectral range of the data.

The pipeline-reduced COS data were downloaded from the {\it HST} archive and processed further using our custom software. The additional data reduction involved recalibrating the COS wavelength solution using a method described in Chen \etal\ (2018) and Zahedy \etal\ (2019). 
These steps resulted in a combined spectrum which was then continuum normalized by fitting a low-order polynomial function to absorption-free spectral regions. The final COS spectrum of HE\,0047$-$1756$A$ has a median signal-to-noise ratio of S/N $\approx 10-20$ per resolution element over the full wavelength range. The wavelength solution is accurate and precise to better than 3 \kms, as evidenced by a comparison between low-ionization absorption features seen in COS and ground-based optical echelle spectra (presented in Z16) and the excellent agreement in line centroids among various $\mathrm{H_2}$ absorption lines spanning $\approx300$ \AA\ in observed wavelength (\S3.1).  

We supplement our COS spectrum of sightline $A$ with low-resolution (${\rm FWHM}\approx270\, \kms $) FUV spectra of both images of the lensed QSO taken with the Space Telescope Imaging Spectrograph (STIS) and the G140L grating onboard {\it HST} from Z17. The STIS spectrum of HE\,0047$-$1756$A$ ($B$) has a median S/N $\approx 20-30\,(12-18)$ per resolution element over its full wavelength range of $1150-1720$ \AA.

\section[]{Results}

\begin{figure}
\begin{center} 
\hspace{-0.16in}
\includegraphics[width=3.2in]{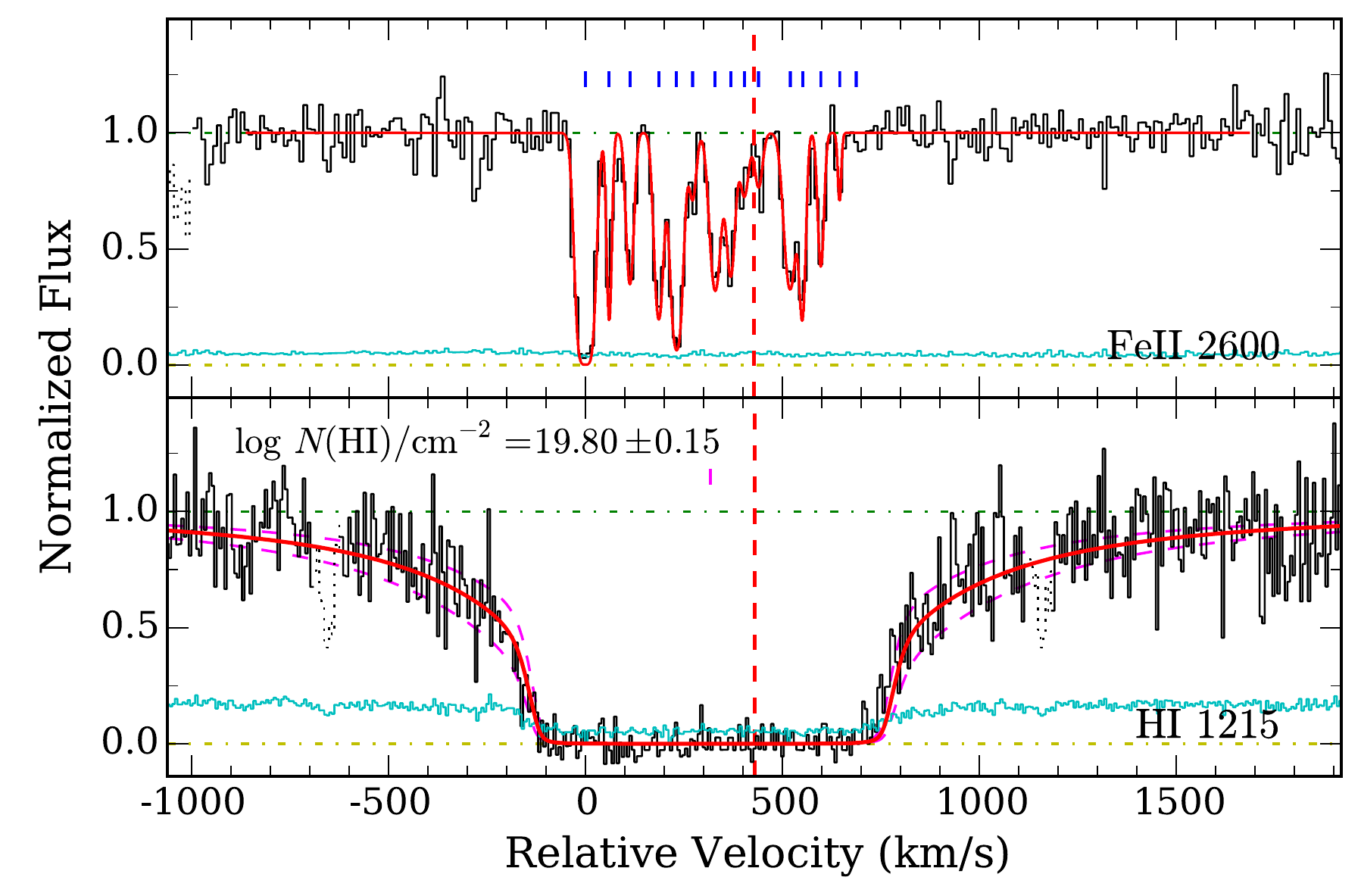}
\end{center}
\vspace{-0.15in}
\caption
{{\it Top}: Kinematically complex gas at $d=4.6$ kpc from the massive elliptical lens galaxy ($z_\mathrm{gal}=0.408$, vertical dashed line), seen in \ion{Fe}{2} $\lambda2600$ absorption from ground-based optical echelle spectrum of HE\,0047$-$1756$A$ (adapted from Z16). The absorption profile comprises 15 individual components (blue tick marks) spanning over $600$ \kms\ in line-of-sight velocity. Zero velocity corresponds to the redshift of the $\mathrm{H_2}$ absorption identified in Figure 2, $z_\mathrm{abs}=0.405985$.  
{\it Bottom}: New {\it HST}/COS FUV spectrum of the corresponding Ly$\alpha$ absorption associated with the lens galaxy. The COS spectrum is rebinned by three pixels for display purposes. The 1-$\sigma$ error spectrum is included in cyan. Contaminating features are dotted out for clarity. The magenta tick mark above the profile indicates the best-fit centroid of the damped Ly$\alpha$ profile. The solid red and dashed magenta curves show the best-fit $N($\ion{H}{1}) and its uncertainty, log $N($\ion{H}{1})$/\cmjj =19.80\pm0.15$.
}
\end{figure}

A prominent feature associated with the massive elliptical lens galaxy is the Ly$\alpha$ absorption with strong damping wings (Figure 1, bottom), confirming the previously reported high $N($\ion{H}{1}) of the gas inferred using low-resolution STIS FUV spectra (Z17). To refine the $N($\ion{H}{1}) measurement, we perform a Voigt profile analysis on the observed damped Ly$\alpha$ profile using a custom software (see Zahedy \etal\ 2019) that takes into account the relevant COS line-spread function (LSF; Lifetime Position 4). Our analysis yields a total \ion{H}{1} column density of log $N($\ion{H}{1})$/\cmjj =19.80\pm0.15$, which is consistent within uncertainties with the Z17 measurement. We adopt this $N($\ion{H}{1}) throughout subsequent analysis.

\subsection{Discovery of $\mathrm{H}_2$ in the ISM of the Lens Galaxy}

A visual inspection of our {\it HST}/COS spectrum of HE\,0047$-$1756$A$ reveals the presence of numerous absorption features consistent with the $\mathrm{H_2}$ Lyman and Werner bands at redshift $z\approx0.406$, or approximately $-430$ \kms\ from the systemic redshift of the lens galaxy (see Figure 2).\footnote{While the velocity offset of the $\mathrm{H_2}$ absorption features may seem large for ISM gas, it is partly explained by the uncertainty on the lens redshift ($\approx 200$ \,\kms ; Z16). Furthermore, the projected escape velocity at $r=5$ kpc from the lens galaxy is $\approx400-500$\,\kms\  given the estimated mass of its host dark-matter halo (Z16), so the observed $\mathrm{H_2}$ kinematics is consistent with ISM gas that is bound to the galaxy. Empirically, large kinematic widths of $\approx500$\,\kms\ have been observed in the atomic/molecular ISM of some nearby early-type galaxies (e.g., Oosterloo \etal\ 2007; Davis \etal\ 2013), reflecting the potential wells of these massive systems.} The $\mathrm{H_2}$ absorption features coincide in velocity with the strongest low-ionization absorption component identified in Z16 (component 1 in their Table 6). We are able to identify more than 140 absorption transitions originating from the ground state of $\mathrm{H_2}$ at different rotational levels from $J=0$ to $J=5$. Each of these transitions has a vibrational quantum number of $\nu=0$ for the lower state and $\nu\leq17$ ($\nu\leq4$) for the upper state in the Lyman (Werner) band.

\begin{figure*}
\begin{center} 
\vspace{0.in}
\includegraphics[width=6.3in]{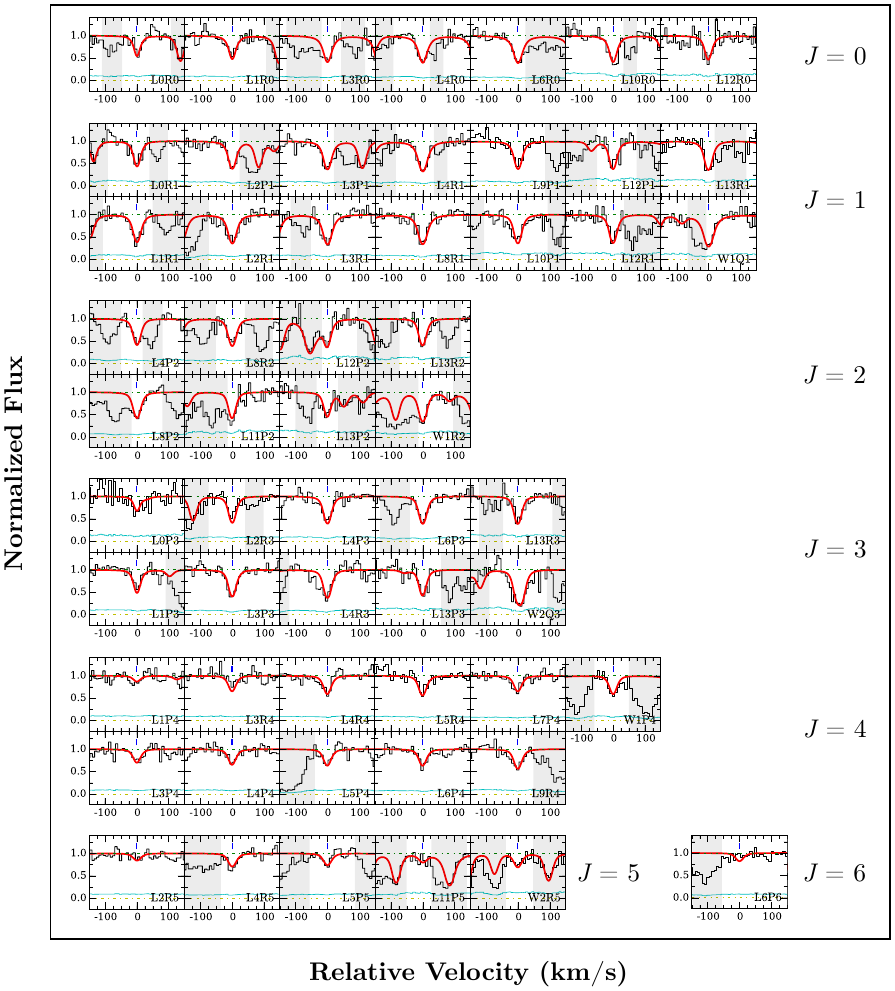}
\end{center}
\vspace{-0.2in}
\caption
{Continuum-normalized absorption profiles of select $\mathrm{H}_2$ Lyman- and Werner-band transitions that are used in our absorption analysis, grouped by $J$ level, observed along sightline HE 0047$-$1756$A$ at $d=4.6$ kpc from the massive elliptical lens. The COS spectrum is rebinned by three pixels for display purposes. The 1-$ \sigma$ error spectrum is included in cyan. Zero velocity marks the best-fit redshift of the $\mathrm{H}_2$ absorption identified with a Voigt profile analysis, $z_\mathrm{abs} = 0.405985$.  Regions excluded from the analysis due to blending and/or contaminating features have been grayed out for clarity. The best-fit $\mathrm{H}_2$ absorption profiles are plotted on top of the data in red curves. The significant detection of $\mathrm{H}_2$ at $J>2$ indicates that  non-thermal excitation mechanism is effective in populating these high rotational levels (see \S 4.1).}
\end{figure*}

To characterize the molecular gas properties, we perform a Voigt profile analysis using a custom software that models the observed $\mathrm{H_2}$ transitions in each $J$ level simultaneously. We adopt the $\mathrm{H_2}$ line list from Ubachs \etal\ (2019) which was made available to us by Patrick Petitjean (private communication). Although we detect more than 140 $\mathrm{H_2}$ transitions, a significant fraction of these lines are blended with each other or other absorption lines. To ensure robust fitting results, we perform our absorption analysis on a subset of available lines (between five and 14 transitions) for each $J$ value, which are selected to contain minimal blending and have unambiguous local continuum level. While only a fraction of observed  $\mathrm{H_2}$ transitions are used to find the best-fit model, we find that the resulting full $\mathrm{H_2}$ absorption model reproduces the absorption profiles of most of the excluded transitions reasonably well.

\begin{table}
\begin{center}
\caption{$\mathrm{H_2}$ properties at $d=4.6$ kpc from the lens galaxy}
\label{tab:Imaging}
\hspace{-0.28in}
\resizebox{2.6in}{!}{
\begin{tabular}{cclc}
\hline
 $z_\mathrm{abs}$		& $J$	&  	 log\,$N/$\cmjj 	& $b$  \\
      					&           	& 				&(\kms)\\
\hline
$0.405985$ 	& 	$0$	&  $17.34^{+0.09}_{-0.26}$ 	&  $2.7^{+0.8}_{-0.5}$   \\
  		    	&	$1$	&  $17.58^{+0.07}_{-0.21}$ 	&  $3.9^{+0.5}_{-0.3}$    \\
     			&	$2$	&  $15.71^{+0.44}_{-0.18}$ 	&  $6.3^{+0.8}_{-1.1}$    \\
     			&	$3$	&  $15.86^{+0.28}_{-0.23}$	&  $6.2^{+1.0}_{-0.8}$    \\
    			&	$4$	&  $14.82^{+0.06}_{-0.04}$	&  $8.7^{+1.6}_{-1.7}$     \\
     			&	$5$	&  $14.61^{+0.06}_{-0.05}$ 	&  $10.5^{+5.0}_{-2.5}$     \\
     			&	$6$	&  $<14.4^a$      			&  $10$     \\   %INCLUDE FOOTNOTE ABOUT CHOICE OF B  ITS EFFECT IF it's changed \hline
     			& $\mathbf{Total}$	&  $17.8^{+0.1}_{-0.3}$      	&      \\  \hline
\multicolumn{4}{l}{$^a$ 95\% upper limit (see \S3.1).}\\
\end{tabular}}
\end{center}
\end{table}

For each rotational $J$ level, we first generate a model spectrum for a single-component line profile, motivated by both the narrow linewidths and lack of kinematic substructures in the observed $\mathrm{H_2}$ absorption profiles (see Figure 2). The Voigt profile is uniquely defined by three free parameters: the line centroid redshift $z_\mathrm{abs}$, the absorption column density $\log\,N$, and the Doppler parameter $b$. To reduce the number of free parameters, all transitions from a given $J$ level are tied to have the same $\log\,N$ and $b$. We further require different $J$ levels to share the same line centroid redshift. Once a theoretical $\mathrm{H_2}$ absorption spectrum has been generated, it is convolved with the relevant COS LSF and subsequently binned to match the pixel resolution of the data. Finally, this model spectrum is compared to the data and the best-fit model parameters for each $J$ level are found by minimizing $\chi^2$ value at the selected $\mathrm{H_2}$ transitions. We estimate the model uncertainties by constructing a marginalized posterior probability distribution for each model parameter based on a Markov Chain Monte Carlo (MCMC) analysis done with the \textsc{Emcee} package (Foreman-Mackey \etal\ 2013). Each MCMC run consists of 500 steps performed by an ensemble of 250 walkers, which are seeded in a small region of the parameter space around the minimum $\chi^2$ solution to speed up convergence. 

We present the best-fit model absorption profiles and compare them to the data in Figure 2. In addition, we summarize the results of the Voigt profile analysis in Table 1, where we report the model parameters and estimated 68\% confidence intervals for the $J=0$ to $J=5$ levels. For $J=6$, which does not exhibit any detectable absorption, we report in Table 1 the 95\% upper limit on the absorption column density for a $b=10$\,\kms\ line profile (matching the linewidth of the $J=5$ level), estimated using the error array at the strongest available $J=6$ transition in the COS data. The best-fit model yields a total $\mathrm{H_2}$ column density of $\log\,N(\mathrm{H_2})/\cmjj=17.8^{+0.1}_{-0.3}$
and a best-fit redshift of $z_\mathrm{abs}=0.405985\pm0.000005$. The centroid of the $\mathrm{H_2}$ line profile is consistent within uncertainties ($< 1\,$\kms) with the strongest low-ionization metal component identified in ground-based optical echelle spectra (Z16), which indicates their association. 

As shown in Table 1, our analysis also identifies a trend of increasing Doppler parameter with increasing $J$ value, from $b\approx3$ \kms\ at $J=0$ to $b\approx10$ \kms\ at $J=5$. The trend of rising velocity dispersion with rotational level has been reported in a number of $\mathrm{H_2}$-bearing damped Ly$\alpha$ absorbers (DLAs) at low and high redshifts (e.g., Ledoux \etal\ 2003; Albornoz V\'asquez \etal\ 2014; Boettcher \etal\ 2020). The kinetic temperature needed to thermally broaden the $\mathrm{H_2}$ line profiles to a linewidth of  $\approx 3\,(10)$ \kms\ is $\approx 10^3\,(10^4)$ K, significantly higher than temperatures at which a significant amount of molecular gas is expected to be present. Therefore, our measurements indicate that non-thermal line broadening is dominant for both low and high $J$ states in the gas, with increasing turbulence toward higher rotational states. We discuss the possible origins of this trend in \S 4.1.

With both the neutral and molecular hydrogen contents of the absorber known, we can estimate the molecular gas fraction according to the following expression,
$f_\mathrm{H_2} = 2\,N(\mathrm{H_2})/[2\,N(\mathrm{H_2})+N(\mathrm{H\,I)}]$.
The extremely strong low-ionization absorber observed along HE\,0047$-$1756$A$ is resolved into 15 kinematic components (Figure 3; Z16). While the total $N$(\ion{H}{1}) can be measured robustly from the strong  Ly$\alpha$ damping wings (Figure 1), it is not possible to constrain the \ion{H}{1} column densities of these individual components because all available \ion{H}{1} Lyman series lines are heavily saturated and the different components blended. Thus, we first estimate $f_\mathrm{H_2}$ by attributing all the observed $N$(\ion{H}{1}) to the $\mathrm{H}_2$-bearing component. Although this assumption is unrealistic because it would result in highly asymmetric  Ly$\alpha$ damping wings owing to the $\mathrm{H_2}$-bearing component occurring at the blue extremum of the profile, it yields a {\it conservative} lower limit on $f_\mathrm{H_2}$ of log\,($f_\mathrm{H_2})_\mathrm{lower}=-1.7^{+0.2}_{-0.3}$. To estimate an upper bound on the molecular gas fraction, we note that the $\mathrm{H}_2$-bearing component contains $\approx40-45$\% of the total column densities of the low-ionization species probed by \ion{Mg}{1}, \ion{Mg}{2}, and \ion{Fe}{2} absorptions (see Table 6 of Z16). If we assume that all 15 components have similar metallicities and dust content, which is justified by the relatively uniform $\mathrm{Fe/Mg}$ elemental abundance ratio observed across all components (Z16), then the inferred $N$(\ion{H}{1}) of the $\mathrm{H}_2$-bearing component is log\,$N($\ion{H}{1})$/\cmjj  \approx19.4$. Consequently, the implied molecular gas fraction is log\,($f_\mathrm{H_2})_\mathrm{upper}=-1.3^{+0.2}_{-0.3}$. 

The observed $f_\mathrm{H_2}$ at $d=4.6$ kpc from the lens galaxy is comparable to nearby ellipticals with CO detections (Welch \etal\ 2010; Young \etal\ 2014) but is
among the highest known for $z<1$ DLAs, where $\approx 90\%$ of absorbers with log\,$N($\ion{H}{1})$/\cmjj \gtrsim 19$ have log\,$f_\mathrm{H_2}\lesssim-2$ (e.g., Crighton \etal\ 2013; Muzahid \etal\ 2015a; 2016; but see Boettcher \etal\ 2020). Considering that $\mathrm{H_2}$ forms on the surface of dust grains, the high $f_\mathrm{H_2}$ can be explained by the high gas metallicity, $\mathrm{[Fe/H]}\gtrsim 0$ (Z17), which results in an elevated dust-to-gas ratio relative to the general DLA population.

\subsection{Highly Ionized Gas in the ISM of the Lens Galaxy}

\begin{figure} 
\begin{center}
\hspace{-0.05in}
\includegraphics[width=3.1in]{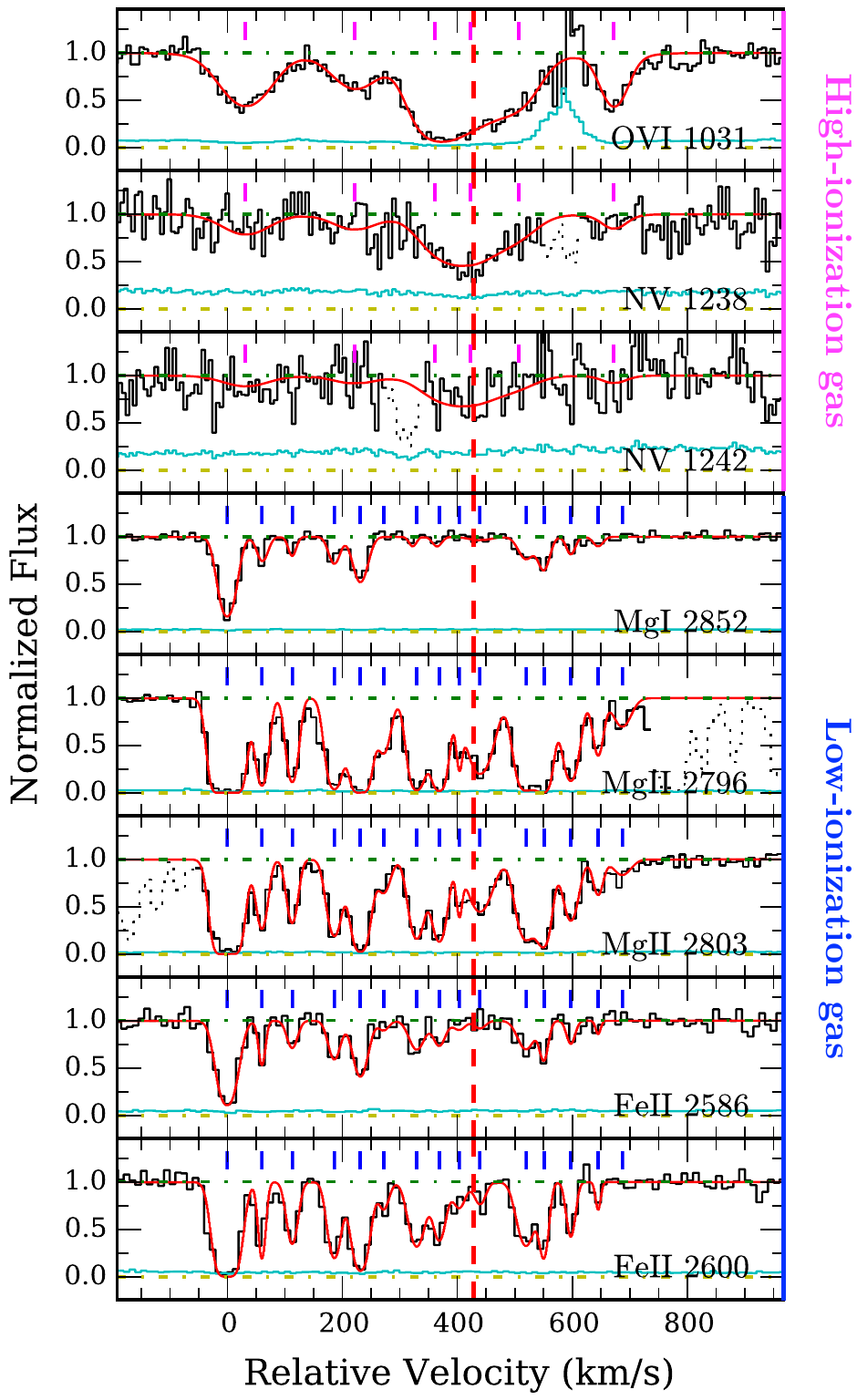}  % CHANGE THIS TO SEPARATE LOW AND HIGH IONS, AND ZERO AT THE MOLECULAR GAS
\end{center}
\vspace{-0.1in}
\caption
  {Continuum normalized absorption profiles of different high- and low-ionization 
  metal transitions along HE\,0047$-$1756$A$ at $d=4.6$ kpc from the massive elliptical lens.  
  Zero velocity corresponds to the redshift of the  $\mathrm{H_2}$ absorption detected in Figure 2, whereas the
  systemic redshift of the lens galaxy, $z_\mathrm{gal}=0.408$, is shown in vertical dashed line.
  The 1-$\sigma$ error spectrum is included in cyan. 
  Contaminating features have been dotted out for clarity.
  The magenta tick marks at the top of the first three panels indicate the
  location of individual components for the high-ionization species identified 
  in the Voigt profile analysis (see \S3.2), with the best-fit Voigt
  profile models included in red. For comparison, individual
  components of the low-ionization species are marked with the blue tick marks in the bottom five panels (Z16). The high ions show a distinct kinematic structure from what is seen in the low ions, indicating that they arise in a different gas phase.}
\label{Figure 7.3}
\end{figure}

\begin{table}
\begin{center}
\caption{High-ionization absorption properties at $d=4.6$ kpc}
\label{tab:Imaging}
\hspace{-0.45in}
\resizebox{3.7in}{!}{
\begin{tabular}{clrrr}\hline

Component	&	Species		&\multicolumn{1}{c}{d${v_c}^a$} 	&\multicolumn{1}{c}{$b$} & \multicolumn{1}{c}{log\,$N$\,/\cmjj}			\\	
 			&				&\multicolumn{1}{c}{(km\,s$^{-1}$)}	& \multicolumn{1}{c}{(km\,s$^{-1}$)}	   &   \\ \hline 
1	& \ion{O}{6}	&	$+32.0\pm3.4$	&$51.4\pm4.9$		& $14.31\pm0.03$	  \\
	& \ion{N}{5}	&				&				& $13.75\pm0.11$	 	\\ \hline	
2	& \ion{O}{6}	&	$+220.1\pm5.6$& $47.9\pm8.9$ 	& $14.02\pm0.06$	\\
	& \ion{N}{5}	&				&				& $13.58\pm0.16$		\\ \hline	
3	& \ion{O}{6}	&	$+372.3\pm6.6$&$42.8\pm3.5$	& $14.71\pm0.06$	  \\
	& \ion{N}{5}	&				&				& $13.76\pm0.15$	 	\\ \hline	
4	& \ion{O}{6}	&	$+423.4\pm14.7$&$64.2\pm21.4$	& $14.60\pm0.13$ \\
	& \ion{N}{5}	&				&				& $14.30\pm0.06$	 	\\ \hline	
5	& \ion{O}{6}	&	$+507.3\pm12.7$&$43.4\pm8.4$ 	& $14.13\pm0.18$	 \\
	& \ion{N}{5}	&				&				& $13.59\pm0.18$	 	\\ \hline	
6	& \ion{O}{6}	&	$+672.3\pm3.0$&$25.5\pm4.6$	& $14.10\pm0.06$	  \\
	& \ion{N}{5}	&				&				& $13.33\pm0.22$	 	\\ \hline

\multicolumn{5}{l}{$^a$ Relative velocity shift from the $\mathrm{H_2}$ absorption redshift, $z_\mathrm{abs}=0.405985$} \\
%\multicolumn{5}{l}{\,\,\,\,at }\\
\end{tabular}}
\end{center}
\end{table}

Z17 previously noted possible absorption features from high-ionization metal lines associated with the lens galaxy along both sightlines of HE\,0047$-$1756. 
However, the low spectral resolution of the Z17 data prevented a detailed investigation into this tentative detection of high-ionization gas.
The {\it HST}/COS spectrum of HE\,0047$-$1756$A$ clearly resolves different metal absorption profiles, enabling precise measurements of gas kinematics and column densities of the highly ionized species.

As shown in the top three panels of Figure 3, the new COS spectrum confirms that \ion{O}{6} absorption is indeed detected and resolved into multiple components in the lens galaxy. In addition, \ion{N}{5} absorption is also detected with a kinematic structure that is consistent with \ion{O}{6}. To constrain their absorption properties, we perform a joint Voigt profile analysis of the \ion{O}{6} $\lambda1031$ line and the  \ion{N}{5} $\lambda\lambda1238,1242$ doublet following the method of Zahedy \etal\ (2019).\footnote{The second member of the  \ion{O}{6} doublet,  \ion{O}{6} $\lambda1037$, is excluded from the Voigt profile analysis due to significant blending with neighboring low-ionization transitions \ion{C}{2} $\lambda1036$ and \ion{O}{1} $\lambda1039$. Although the \ion{O}{6} $\lambda1031$ profile is also contaminated by a higher-redshift Ly$\epsilon$ line at $z=0.55003$, in this case the absorption profile of the contaminating Ly$\epsilon$ line is well-constrained by various other Lyman series lines observed in our COS spectrum. To remove this contamination from the \ion{O}{6} $\lambda1031$ absorption, we have divided the observed \ion{O}{6} $\lambda1031$ profile by the best-fit model of the $z=0.55003$ Ly$\epsilon$ line prior to performing the analysis.} To ensure the robustness of the fit, we tie both the component structure and Doppler linewidths of the two ions. We summarize the results from our Voigt profile analysis of these high-ionization species in Table 2. The continuum-normalized absorption profiles and best-fit models for \ion{O}{6} and \ion{N}{5} are presented in the top three panels of Figure 3. To compare the kinematics between low- and high-ionization species, we also show the observed and modeled absorption profiles of \ion{Mg}{1}, \ion{Mg}{2}, and \ion{Fe}{2} in the bottom five panels of Figure 3, from previous absorption analysis reported in Z16. 

It is clear from Figure 3 that the high ions exhibit a distinct kinematic structure from that of the low ions, which indicates that the high ions arise in a different gas phase (Zahedy \etal\ 2019). Specifically, our analysis reveals a highly ionized gas phase in the lens ISM that is kinematically complex, comprising six broad kinematic components ($b\approx25-65\, \kms$) that span $\approx 640$ \kms\ in line-of-sight velocity. The observed total column densities of these highly ionized species are log $N$(\ion{O}{6})$/\cmjj =15.2\pm0.1$ and log $N$(\ion{N}{5})$/\cmjj\ =14.6\pm0.1$. These \ion{O}{6} and \ion{N}{5} absorbers are among the strongest known to be in the vicinity of $z<1$ galaxies (cf., Johnson \etal\ 2015; Muzahid \etal\ 2015b; Werk \etal\ 2016; Rosenwasser \etal\ 2018; Zahedy \etal\ 2019), where high-ionization absorbers with log $N$(\ion{O}{6})$/\cmjj > 15$ and log $N$(\ion{N}{5})$/\cmjj > 14$ are rare.   

It is also interesting to note the observed \ion{N}{5} to \ion{O}{6} column density ratios among the six high-ionization components, which have an arithmetic mean and dispersion of log\,$\langle N$(\ion{N}{5})$/N$(\ion{O}{6}) $\rangle=-0.5\pm0.2$. These ionic ratios are considerably higher than typical values seen in the Galactic corona (e.g., Wakker \etal\ 2012), the circumgalactic medium of external galaxies (e.g., Werk \etal\ 2016; Zahedy \etal\ 2019), and the high-redshift intergalactic medium (e.g., Lehner \etal\ 2014), where a large majority of absorbers in these diverse environments exhibit log\,$N$(\ion{N}{5})$/N$(\ion{O}{6}) $\lesssim-0.8$.  We argue that a super-solar $\mathrm{[N/O]}$ in the highly ionized gas phase is the most likely explanation, considering that high nitrogen-to-alpha ratios of  $\mathrm{[N/\alpha]}\gtrsim 0.3$ have been reported in the evolved stellar populations and cool ISM of nearby ellipticals (e.g., Greene \etal\ 2013; Yan 2018). Similar $\mathrm{[N/O]}$ ratios in both high- and low-ionization gases would also suggest a causal link between different phases of the ISM of the elliptical lens (we discuss this connection in \S 4.3).

\begin{figure*}
\begin{center} 
\hspace{-0.2in}
\includegraphics[width=7.2in]{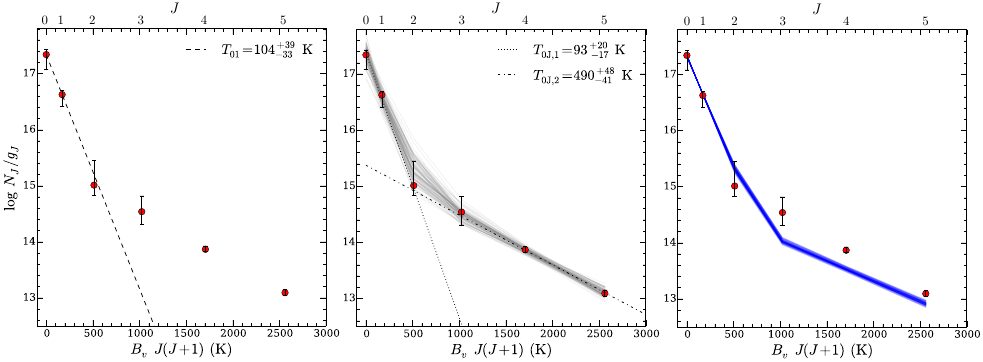}
\vspace{-0.2in}
\end{center}
\caption
{Excitation diagram for the observed rotational level populations  (red circles) of $\mathrm{H_2}$ gas at $d=4.6$ kpc from the massive elliptical lens. {\it Left}: Assuming that different rotational levels follow a Boltzmann distribution, the observed ratio between the $J=0$ and $J=1$ levels indicates an excitation temperature of  $T_\mathrm{01}=104^{+39}_{-33}$ K (dashed line). However, this single-temperature model severely underpredicts the observations at $J>2$. {\it Middle}: A model with two excitation temperatures of $T_\mathrm{0J,1}=93^{+20}_{-17}$ K (dotted line) and $T_\mathrm{0J,2}=490^{+48}_{-41}$ K (dash-dotted line) can reproduce the observed level populations. The thin gray curves show 100 random realizations of the two-temperature model using the MCMC method. {\it Right}: The thin blue curves represent a set the \textsc{cloudy} models that best reproduce the trend seen in the data. These models have UV radiation fields which are $15-25$ times more intense than the Milky Way ISM radiation field. If the elevated populations at higher rotational levels are due to radiation pumping, the required radiation field is significantly higher than what is observed in the local Galactic ISM. }
\end{figure*}

\section[]{Discussion}

\subsection{Physical Conditions of the $\mathrm{H}_2$ Gas}

The distribution of $\mathrm{H_2}$ molecules among different rotational levels reflects the excitation state of the gas and offers insight into 
the physical mechanisms that are responsible. The relative populations of different $\mathrm{H_2}$ rotational levels can be described by a Boltzmann distribution, 
$\frac{N_J}{N_{J=0}}  = \frac{g_J}{g_{J=0}}\,\mathrm{exp}[-{B_v\,J(J+1)}/{T_\mathrm{0J}}]$,
where $N_\mathrm{J}$ is the $\mathrm{H_2}$ column density for the rotational level $J$, $T_\mathrm{0J}$ is the excitation temperature from $J=0$ to rotational level $J$, and $B_v = 85.36$ K. The statistical weight $g_J$ is $(2J+1)$ for even-numbered $J$ or $3(2J+ 1)$ for odd-numbered $J$. In Figure 4, we show the $\mathrm{H}_2$ excitation diagram of the absorber for rotational states between $J=0$ and $J=5$. The column density ratio between $J=0$ and 1 states, which contain $\approx 98$\% of the total $N(\mathrm{H_2})$, implies an excitation temperature of $T_\mathrm{01}=104^{+39}_{-33}$ K. 

The observed $T_\mathrm{01}$ for the bulk of molecular gas along the lensed sightline is comparable to typical values reported in $z<1$ $\mathrm{H}_2$-bearing DLAs (see Muzahid \etal\ 2015a). However, while the observed population for $J=2$ can also be well-reproduced by the same excitation temperature, this single-temperature model fails to explain the observed column densities at $J>2$ (Figure 4, left panel). The predicted column densities for these higher rotational states are orders of magnitude lower than the observations, which indicates a higher excitation temperature for $J>2$. 

It is well-known from Galactic $\mathrm{H}_2$ studies that a one-temperature fit typically works only for optically thin $\mathrm{H}_2$ absorbers with log\,$N(\mathrm{H_2})/ \cmjj \lesssim 15 $ (e.g., Spitzer \etal\ 1974; Spitzer \& Jenkins 1975; Jenkins \& Peimbert 1997). In contrast, stronger $\mathrm{H_2}$ absorbers in the Galaxy and beyond have been found to exhibit elevated populations at higher rotational levels  (e.g., Jenkins \& Peimbert 1997; Reimers \etal\ 2003; Noterdaeme \etal\ 2007; Rawlins \etal\ 2018; Balashev \etal\ 2019; Boettcher \etal\ 2020), which indicate that the $\mathrm{H}_2$ gas is bifurcated into two excitation temperatures. Motivated by these prior observations, we perform a simultaneous fit of a two-temperature model to our data and find that the observed $\mathrm{H}_2$ level populations are well-reproduced by two excitation temperatures of $T_\mathrm{0J,1}=93^{+20}_{-17}$ K and $T_\mathrm{0J,2}=490^{+48}_{-41}$ K (Figure 4, middle panel).

The observed temperature bifurcation and trend of rising velocity dispersion with $J$ level (see \S3.1) can be understood to be a consequence of the $\mathrm{H}_2$ absorption originating in a gas cloud with an internal density and/or temperature stratifications (e.g., Noterdaeme \etal\ 2007). In this scenario, most of the column densities at low-$J$ levels originate from the inner layer of the cloud, where the gas is sufficiently dense and shielded from radiation that collisions are the dominant excitation mechanism. Consequently, the low-level populations are essentially thermalized and the lower excitation temperature is highly coupled to the kinetic temperature of the gas.

In contrast, the elevated column densities and broader line profiles of high-$J$ levels indicate that they arise primarily from warmer and more turbulent outer layers of the cloud (e.g., Lacour \etal\ 2005). At these locations, $\mathrm{H_2}$ molecules can be highly excited through collisions triggered by shocks and turbulent dissipation (e.g. Jenkins \& Peimbert 1997; Gry \etal\ 2002; Gredel \etal\ 2002; Ingalls \etal\ 2011), as well as through radiation pumping by an external UV radiation field (e.g., Jura 1975; Klimenko \& Balashev 2020). A unique prediction of the shock scenario is a systematic shift of up to a few \kms\ in line centroids with increasing rotational state, which is caused by the different $J$ levels originating from distinct locations moving at slightly different speeds relative to the shock front (e.g. Jenkins \& Peimbert 1997; Gredel \etal\ 2002). Although we do not detect any systematic shift in line centroids with $J$ levels to within the precision of our COS wavelength calibration ($\lesssim 3$ \kms), we cannot rule out a more modest shift of $\approx 1$ \kms\ or less, which may be the result of weaker shocks (e.g., Gredel 1997). 

As an alternative, we now explore radiation pumping as an excitation mechanism. We perform a series of calculations using the \textsc{Cloudy} v.13.03 code (Ferland \etal\ 2013) to simulate a plane-parallel slab of gas with uniform density $n_\mathrm{H}$ which is irradiated by two UV radiation fields: the updated Haardt \& Madau (2001) extragalactic UV background at $z=0.4$, known as HM05 in \textsc{Cloudy}, and the built-in unextinguished Milky Way ISM radiation field from Black (1987). To constrain the strength of UV radiation that is required to reproduce the observations, we vary the overall intensity of the ISM radiation field by a scale factor of between 0.1 and 100. We incorporate dust grains in the calculations following the observed grain abundance and size distribution in the local ISM. For each input radiation field, we construct a grid of \textsc{Cloudy} models spanning a wide range of gas densities ($0\leq \mathrm{log}\,n\mathrm{_H/cm^{-3}}\leq 4$) at the observed gas metallicity (Z17). For each grid point, \textsc{Cloudy} calculates the expected column density for each $J$ level assuming thermal and ionization equilibrium. To simulate two-sided illumination of the cloud, we use half the observed $N(\mathrm{H_2})$ as the stopping condition for the calculations and subsequently double the output $\mathrm{H_2}$ level populations for comparison with the data. 

We summarize the results of our \textsc{Cloudy} calculations in the right panel of Figure 4, where the set of models that best reproduce the observed $\mathrm{H_2}$ excitation diagram are shown in thin blue curves. These models have UV radiation fields which are $15-25$ times stronger than the local ISM radiation field. The range of gas densities are $n_\mathrm{H}\approx1000-2500\,\cmjjj$, with mean $\mathrm{H_2}$ kinetic temperatures ($90-130$ K) and \ion{H}{1} column densities (log $N($\ion{H}{1})$/\cmjj =19.6-19.9$) which are broadly consistent with the observations. While it is clear that these simple models are only able to roughly reproduce the general trend seen at $J>2$, this exercise demonstrates that if the elevated populations at higher rotational levels are primarily due to radiation pumping, the required UV radiation field is significantly higher than what is observed in the Galactic ISM (see also e.g., Klimenko \& Balashev 2020; Boettcher \etal\ 2020).

\subsection{Spatial Variations in Multiphase Gas Properties}

A benefit of using a multiply lensed QSO system as gas probes is the ability to investigate spatial variations in the gas properties of a foreground galaxy. As described in Z16 (see their Figure 1), the doubly lensed images of HE\,0047$-$1756 probe opposite sides of the massive elliptical lens galaxy, with sightline $A$ at $d=4.6$ kpc (1.8 half-light radii, $r_e$) and sightline $B$ at $d = 3.3$ kpc (1.3 $r_e$). The observed \ion{H}{1} and low-ionization metal column densities differ by less than $0.1-0.2$ dex between the two sightlines (Z16; Z17), despite a separation of $\approx 8$ kpc in projection. These similarities suggest that the cool ($T\sim10^4$ K) ISM phase is spatially extended, with a high gas covering fraction at $d\lesssim5$ kpc.

While we are unable to perform a detailed analysis on the \ion{O}{6} absorption detected in the low-resolution STIS FUV spectrum of HE\,0047$-$1756$B$ (Z17), we can compare the general absorption properties of the  \ion{O}{6} absorbers detected along the two sightlines. Specifically, the total  \ion{O}{6} rest-frame equivalent width is $W_r (1031)_B = 1.3\pm0.1$ \AA\ along sightline $B$, which is very similar to what is observed in the COS spectrum along sightline $A$, $W_r (1031)_A = 1.14\pm0.04$ \AA. Furthermore, the observed FWHM of the \ion{O}{6} profile along sightline $B$ is $\approx 580$ \kms, which is comparable to the observed kinematic spread of $\approx 640$ \kms\ for the  \ion{O}{6} absorption profile along sightline $A$ (\S3.2). The coherent \ion{O}{6} absorption properties between the two sightlines imply that similar to the low-ionization gas phase, the highly ionized ISM is spatially extended and has a high covering fraction on a scale of $\sim5$ kpc in the massive elliptical.

The lack of a high-resolution FUV spectrum of HE\,0047$-$1756$B$ prevents a direct search for $\mathrm{H_2}$ along this sightline. To assess whether we can constrain spatial variations in molecular gas properties using the available low-resolution STIS FUV spectrum of sightline $B$, we perform the following experiment. First, we divide the best-fit model for the full Lyman and Werner bands from the high-resolution COS FUV spectrum of sightline $A$ to remove all $\mathrm{H_2}$ absorption from the spectrum. Then, we convolve the resulting ``$\mathrm{H_2}$-free'' spectrum of the QSO with the STIS LSF and compare the result with our STIS spectrum of sightline $A$. We find that while individual $\mathrm{H_2}$ lines are unresolved in the STIS spectrum, the combined absorptions from the Lyman and Werner bands result in an overall flux decrement that is detectable across the QSO spectrum. 

Motivated by the result of the experiment, we generate a series of model Lyman and Werner bands spanning a wide range of $N(\mathrm{H_2})$, apply them to the ``$\mathrm{H_2}$-free'' QSO spectrum, and convolve the results with the STIS LSF. Each of the resulting spectra is then rescaled to the level of sightline $B$ using the mean observed flux ratio of the two lensed images in two absorption-free regions: $1575-1585$ and $1640-1650$ \AA\ in the observed frame. Finally, we compare the products to the STIS spectrum of sightline $B$ in the spectral region between 1415 and 1435 \AA\ in observed wavelength, which has a large concentration of strong $\mathrm{H_2}$ transitions, and infer the allowed $N(\mathrm{H_2})$ using a $\chi^2$ analysis. The observed spectrum of HE\,0047$-$1756$B$ is consistent with the presence of $\mathrm{H_2}$ with $N(\mathrm{H_2})\lesssim10^{16}\,$\cmjj\ at the 95\% confidence level. 

The inferred molecular gas fraction of  $f_\mathrm{H_2}\lesssim0.05\%$ along sightline $B$ is a factor of at least $\approx 40-100$ times lower than that observed along sightline $A$ on the opposite side of the galaxy, $f_\mathrm{H_2}=2-5\%$ (\S 3.1). This exercise suggests that in contrast to the neutral and highly ionized gas phases, the molecular gas distribution in the lens ISM is clumpier. Furthermore, the observed $f_\mathrm{H_2}$ along the two lensed sightlines are consistent with nearby quiescent galaxies found to harbor molecular gas (e.g., Young \etal\ 2014) but low compared to typical values in star-forming disks (Chen 2017b and references therein). If these $f_\mathrm{H_2}$ constraints are representative of the rest of the galaxy, they imply a low mass fraction of dense, cold molecular gas in the multiphase ISM of the lens. 

\subsection{Implications for Feedback in Massive Ellipticals}

How the ISM is partitioned by mass into its different gas phases depends sensitively on the gas cooling rate, the available heating to offset this cooling, and the relevant timescales of these processes. The simultaneous detections of multiple gas phases in the lens ISM enable such an investigation for the first time in a distant elliptical, which can offer valuable insight into late-time feedback in massive elliptical galaxies.

Specifically, now that we have robustly detected highly ionized gas in the lens galaxy and constrained its properties, we can calculate the mass budget in the warm ($T\sim10^5$ K) gas phase and compare it to the previously estimated mass budget in the cool ISM (Z17). For the cool phase, Z17 estimated a total Fe mass of $M_\mathrm{Fe}\sim (5-8)\times10^4 \,(f_\mathrm{c,cool}) \,\mathrm{M_\odot}$ at $d<5$ kpc ($\approx 2\,r_e$, matching the region probed by the doubly lensed QSO), where $f_\mathrm{c,cool}$ is the cool gas covering fraction. The corresponding total mass in the cool phase is 
%$M_\mathrm{cool}\sim (4-6)\times10^7 \, \,(f_\mathrm{cov}\, Z) \,\mathrm{M_\odot}$
\begin{equation}
M_\mathrm{cool}\sim (4-6)\times10^7  \bigg(\frac{f_\mathrm{c,cool}}{1.0}\bigg)  \bigg(\frac{Z_\mathrm{cool}}{Z_\odot}\bigg)^{-1} \, \mathrm{M_\odot},
%\frac{N_J}{N_{J=0}} = \frac{g_J}{g_{J=0}}\,\mathrm{exp}\bigg(-\frac{B_v\,J(J+1)}{T_\mathrm{0J}}\bigg). 
\end{equation}
where $Z_\mathrm{cool}$ is the cool gas metallicity. 
For $f_\mathrm{c,cool}\approx1$ and a solar metallicity gas, which Z17 inferred for the cool phase, the inferred mass in the cool ($T\sim10^4$ K) ISM is $M_\mathrm{cool}\sim (4-6)\times10^7 \, \,\mathrm{M_\odot}$.

Assuming that the observed $N$(\ion{O}{6}) along sightline $A$ is representative at $d<5$ kpc, the estimated mass in the \ion{O}{6}-bearing phase of the ISM is
\begin{equation}
M_\mathrm{warm}\sim 3\times10^7  \bigg(\frac{f_\mathrm{c,warm}}{1.0}\bigg)  \bigg(\frac{Z_\mathrm{warm}}{Z_\odot}\bigg)^{-1} \bigg(\frac{f_\mathrm{O^{5+}}}{0.1}\bigg)^{-1} \mathrm{M_\odot},
%\frac{N_J}{N_{J=0}} = \frac{g_J}{g_{J=0}}\,\mathrm{exp}\bigg(-\frac{B_v\,J(J+1)}{T_\mathrm{0J}}\bigg). 
\end{equation}
where $f_\mathrm{c,warm}$ is the covering fraction of the warm phase, $Z_\mathrm{warm}$ is its metallicity, and $f_\mathrm{O^{5+}}$ is the ionization fraction of $\mathrm{O^{5+}}$ ions. If we further assume a unity covering fraction and solar gas metallicity for the warm ($T\sim10^5$ K) ISM phase, and adopt a reasonable $f_\mathrm{O^{5+}}\approx0.1-0.2$ which is predicted for a wide range of physical conditions at $T\sim10^5$ K (e.g., Oppenheimer \& Schaye 2013), we find a total gas mass of $M_\mathrm{warm}\sim (1.5-3)\times10^7 \, \,\mathrm{M_\odot}$ in the warm  ISM phase that is likely traced by \ion{O}{6} absorption. Despite the uncertainties inherent in our simple calculations, the estimated mass budgets in the cool and warm ISM phases are comparable to within a factor of a few if the two phases have similar gas covering fractions and metallicities. 

In the physical picture where the observed \ion{O}{6} absorber traces transitional temperature ($T\sim10^5$ K) gas that is radiatively cooling from a virialized hot phase ($T\sim10^6$\,K), $M_\mathrm{{warm}}$ is proportional to the mass flow rate into the cool ISM following $\dot{M}_\mathrm{cool}=M_\mathrm{warm}/t_\mathrm{cool}$,  where $t_\mathrm{cool}$ is the cooling timescale of the \ion{O}{6}-bearing gas. The cooling timescale depends on the gas temperature, metallicity, and density. For $T\approx10^{5.5}$ K and a solar-metallicity gas with a density of $n_\mathrm{H}=10^{-3}\,\cmjjj$, typical of the hot halo of massive ellipticals at $d\approx 10$ kpc (e.g., Singh \etal\ 2018), the expected cooling time is $t_\mathrm{cool}\approx 20-30$ Myr (Gnat \& Sternberg 2007; Oppenheimer \& Schaye 2013) with a total cooling rate of $\sim(1-3)\times10^{47}\,\mathrm{erg\,yr^{-1}}$. Thus, in a radiative cooling scenario the estimated $M_\mathrm{{warm}}$ translates to a mass flow rate of $\dot{M}_\mathrm{cool}\sim 0.5-1.5\,\mathrm{M_\odot\,yr^{-1}}$ at $d<5$ kpc. If the bulk of this flow cooled to  $T\lesssim10^4$ K and remained in this phase, we should expect $M_\mathrm{cool} \gg M_\mathrm{warm}$ over a timescale of $\sim100$ Myr in the absence of significant star-formation activity. Considering Z16 found a minimum stellar population age of $>1$ Gyr and no detectable star formation ($\mathrm{SFR}<0.1\,\mathrm{M_\odot\,yr^{-1}}$) in the lens galaxy, this calculation suggests that most of the cooling gas is reheated to the coronal phase.\footnote{In principle, the cool gas could also be depleted primarily by further cooling into the cold ($T\lesssim10^2$ K) phase probed by $\mathrm{H_2}$ molecules. However, we consider this scenario unlikely given the inferred low mass fraction of molecular gas in the lens ISM, which is also consistent with observations of nearby ellipticals (\S 4.2).} To heat the gas back to virial temperature,  the required heating rate is $\dot{E}_\mathrm{heat}\sim(1-3)\times10^{48}\,\mathrm{erg\,yr^{-1}}$, assuming $T_\mathrm{vir}\approx3\times10^6$\,K given the estimated mass of the dark-matter host halo of the lens (Z16).%\footnote{$T_\mathrm{vir}$ is estimated assuming an isothermal gas for the estimated dark-matter host halo mass of the lens galaxy in Z16.}

Observations of nearby massive ellipticals show that mechanical feedback (often dubbed ``radio-mode feedback'') from active galactic nuclei (AGNs) can output as much power as $\dot{E}_\mathrm{AGN}\sim10^{49}-10^{50}\,\mathrm{erg\,yr^{-1}}$ (e.g., Werner \etal\ 2019). If the lens galaxy of HE\,0047$-$1756 hosts an active nucleus at present, then in principle it has more power than what is required to reheat the \ion{O}{6}-traced cooling gas. There are two caveats to this statement, however. Because the estimated cooling time of a $T\sim10^5$ K gas is short ($\sim10^7$ yr) owing to the expected high gas densities at $d<10$ kpc, the actual amount of available heating depends sensitively on the radio-mode duty cycle (i.e., the fraction of time that an AGN is in radio mode). The radio-mode duty cycle in ellipticals has been estimated to be no more than $\approx 30\%$ outside of rich cluster environments (O'Sullivan \etal\ 2017). Furthermore, even if an AGN is currently on, its energy output is likely to be distributed over a large volume in the gaseous halo. Indeed, observations of large X-ray cavities/bubbles and extended radio lobes around nearby giant elliptical galaxies indicate that AGNs deposit kinetic energy on scales of $\sim 50$ kpc or larger in the hot halo (e.g., McNamara \& Nulsen 2007; Fabian 2012). In conclusion, whether AGNs are viable as a continuous heating source requires not only a high duty cycle but also that its mechanical energy can be effectively coupled with the ISM on $\sim1$ kpc scales in the galaxy. 
 
Alternatively, we consider heating sources associated with the old stellar populations themselves. Previous analytic and simulation studies suggest that 
feedback from SNe Ia and stellar winds from asymptotic giant branch (AGB) stars may offset radiative cooling from diffuse gas in massive elliptical galaxies with $M_{\rm star}\approx 10^{11}\, \mathrm{M_\odot}$  (e.g., Conroy \etal\ 2015; Li \etal\ 2018). Empirically, the observed high $\mathrm{[Fe/Mg]}$ abundance ratios at $d\lesssim20$ kpc from quiescent galaxies (Z16; Zahedy \etal\ 2017a) also supports the idea that their ISM has been subjected to significant influence from recent SNe Ia. Using the mean SN Ia rate in nearby ellipticals (e.g., Mannucci \etal\ 2005), Z17 estimated an integrated SN Ia rate of $\sim0.3$ per century within $d<5$ kpc from the massive elliptical lens galaxy. Multiplying this rate by a mean energy of $10^{51}\,\mathrm{erg}$ per SN Ia,  we estimate that the heating rate available from SNe Ia is $\dot{E}_\mathrm{Ia}\sim3\times10^{48}\,\mathrm{erg\,yr^{-1}}$, which is comparable to the required heating. 

In addition to SNe Ia heating, Conroy \etal\ (2015) also considered how materials ejected from AGB stars can interact with and heat the ambient ISM in elliptical galaxies. Using their analytic formula for AGB heating rate, we estimate a heating rate of $\dot{E}_\mathrm{AGB}\sim5\times10^{47}\,\mathrm{erg\,yr^{-1}}$ in the lens galaxy of HE\,0047$-$1756.  This exercise suggests that heating from SNe Ia and AGB stars may suffice to match the cooling rate inferred from the observed \ion{O}{6} absorption and prevent a large accumulation of cold gas in the ISM, even in the absence of strong feedback from an active nucleus. 

\section[]{Conclusions}

Our analysis of the medium-resolution FUV spectrum of lensed QSO sightline HE\,0047$-$1756$A$ has revealed a complex, multiphase gas at $d=4.6$ kpc from the lens and yielded the first constraints on multiphase ISM properties in a massive quiescent galaxy ($M_{\rm star}\approx 10^{11}\, \mathrm{M_\odot}$) beyond the local Universe. $\mathrm{H_2}$ gas is detected with column density $\log\,N(\mathrm{H_2})/\cmjj=17.8^{+0.1}_{-0.3}$ and a molecular gas fraction of  $f_\mathrm{H_2}=2-5\%$. Furthermore, the ISM exhibits \ion{O}{6} and \ion{N}{5} absorptions with a distinct kinematic structure from that of the low ions (e.g. \ion{Mg}{2}; Z16), indicating that these high ions arise in a different gas phase. The highly ionized phase has a total log $N$(\ion{O}{6})$/\cmjj =15.2\pm0.1$ and log $N$(\ion{N}{5})$/\cmjj\ =14.6\pm0.1$, among the strongest associated with $z<1$ galaxies. The low- and high-ionization gas phases are spatially extended on $\sim5$ kpc scale, which is in contrast to the patchier $\mathrm{H_2}$ spatial distribution on this scale. 

We have investigated how the ISM is partitioned by mass into its different phases and examined its implications on late-time feedback in the galaxy. Specifically, the mass in the highly ionized ISM phase is $M_\mathrm{warm}\sim (1.5-3)\times10^7 \, \,\mathrm{M_\odot}$ at $d<5$ kpc, comparable to the estimated mass in the cool ($T\lesssim10^4$ K) ISM. Assuming the high-ionization gas originates in a transient warm ($T\sim10^5$ K) phase undergoing radiative cooling from a hot halo surrounding the galaxy, the inferred mass accretion rate is $\sim 0.5-1.5\,\mathrm{M_\odot\,yr^{-1}}$. The lack of star-formation activity ($\mathrm{SFR}<0.1\,\mathrm{M_\odot\,yr^{-1}}$) in the galaxy suggests that most of this flow is reheated to the hot phase, at a rate of $\dot{E}_\mathrm{heat}\sim(1-3)\times10^{48}\,\mathrm{erg\,yr^{-1}}$. Continuous heating from evolved stellar populations (primarily SNe Ia but also AGB winds) in the massive elliptical galaxy may suffice to prevent a large accumulation of cold gas in the ISM, even in the absence of strong AGN feedback. While this conclusion is based on a single galaxy, our study underscores the important role that evolved stellar populations can play in maintaining the low star-formation rate in massive quiescent galaxies over cosmic time.

\acknowledgements

The authors thank the anonymous referee for thoughtful comments that helped improve the presentation of this paper. 
We thank Patrick Petitjean for providing his $\mathrm{H_2}$ line list, and Sean Johnson and Ben Rosenwasser for insightful discussions. 
FSZ acknowledges support of a Carnegie Fellowship from the Observatories of the Carnegie Institution for Science. FSZ and HWC acknowledge partial support from HST-GO-15250.004A. FSZ, HWC, and  EB acknowledge partial support from HST-GO-15163.001A and NSF AST-1715692 grants.
This work is based on data gathered with the NASA/ESA
{\it Hubble Space Telescope} operated by the Space Telescope Science Institute and the Association of Universities for Research in
Astronomy, Inc., under NASA contract NAS 5-26555. Additional data shown here were gathered with the 6.5 m Magellan Telescopes located at Las Campanas Observatory in Chile.

\end{document}